# Wafer-Scale, Sub-5 nm Junction Formation by Monolayer Doping and Conventional Spike Annealing

Johnny C. Ho,[1,2,3] Roie Yerushalmi,[1,2,3,†] Gregory Smith,[4] Prashant Majhi,[4] Joseph Bennett,[5] Jeffri Halim,[6] Vladimir N. Faifer, [6]Ali Javey[1,2,3,*]

[1]Department of Electrical Engineering and Computer Sciences, University of California at Berkeley, Berkeley, CA, 94720, USA.

[2]Materials Sciences Division, Lawrence Berkeley National Laboratory, Berkeley, CA 94720, USA.

[3]Berkeley Sensor and Actuator Center, University of California at Berkeley, Berkeley, CA, 94720, USA.

[4]SEMATECH, Austin, TX, 78741, USA.

[5]SVTC Technologies, Austin, TX, 78741, USA.

[6]Frontier Semiconductor, Inc., San Jose, CA, 95112, USA.

[†] Current address: Institute of Chemistry, The Hebrew University of Jerusalem

* Corresponding author: ajavey@eecs.berkeley.edu

**ABSTRACT -** We report the formation of sub-5 nm ultrashallow junctions in 4” Si wafers enabled by the molecular monolayer doping of phosphorous and boron atoms and the use of conventional spike annealing. The junctions are characterized by secondary ion mass spectrometry and non-contact sheet resistance measurements. It is found that the majority (~70%) of the incorporated dopants are electrically active, therefore, enabling a low sheet resistance for a given dopant areal dose. The wafer-scale uniformity is investigated and found to be limited by the temperature homogeneity of the spike anneal tool used in the experiments. Notably, minimal junction leakage currents (<1 μA/cm$^2$) are observed which highlights the quality of the junctions formed by this process. The results clearly demonstrate the versatility and potency of the monolayer doping approach for enabling controlled, molecular-scale ultrashallow junction formation without introducing defects in the semiconductor.

Device scaling has been the main driving force for the technology advancement in the semiconductor industry over the last few decades.[1,2] Specifically, junction depths have been scaled continuously together with the gate lengths in order to achieve faster transistor speeds and higher packing densities. Historically, source/drain extension junction depths of ~1/3 of the transistor gate lengths have been used for efficient electrostatics and acceptable leakage currents. With the gate lengths fast approaching the sub-10 nm regimes, it is vital to realize sub-5 nm ultrashallow junctions (USJs) with low sheet resistivity (i.e., low parasitic contact resistance) to facilitate the future scaling of transistors. However, there are tremendous technological challenges for achieving sub-5 nm USJs as the conventional doping strategies suffer from a number of setbacks.[1,2] The current USJs are fabricated by the combination of ion implantation and spike annealing. During the process, Si atoms are displaced by energetic dopant ions and a subsequent annealing step (e.g., spike, a high temperature anneal process of less than 1s with fast temperature ramp up/down capability) is used to activate the dopants by moving them into the appropriate lattice positions and restoring the substrate's crystal quality. However, point defects such as Si interstitials and vacancies are also generated, which interact with the dopants to further broaden the junction profile. This is known as the transient enhanced diffusion (TED)[3] which limits the formation of sub-10nm USJs by conventional technologies. Moreover, there are significant research efforts to develop new strategies such as utilizing heavier implantation dopant sources (molecular implantation[4], gas cluster ion beam[5] and plasma doping[6]) to obtain shallower doping profiles, and advanced annealing techniques (flash[7] and laser[8]) to activate the implanted dopants without causing significant diffusion. However, very little is known on the effects of advanced doping and annealing techniques on the junction uniformity, reliability and subsequent process integration which may hamper their use in the IC manufacturing. To address the need for advanced doping strategies, recently, we reported a new technique for controlled, nanoscale surface doping of semiconductor materials by utilizing the crystalline nature of silicon and its self-limiting surface reaction properties to form self-assembled dopant monolayers followed by a subsequent annealing step for the incorporation and diffusion of dopants.[9] Due to the lack of damage to the lattice during this surface doping strategy, minimal TED effects are expected which presents a major advantage of this technology for achieving nanoscale junctions. Previously, we reported junction depths down to ~20 nm by utilizing the monolayer doping method and 5 sec rapid thermal diffusion.[9] Here, we explore and characterize the junction depth limits of this process on wafer-scale and in detail through materials and electrical characterizations. Uniquely, for the first time, we demonstrate sub-5 nm junction depths (down to ~2nm; at the limit of most characterization methods) with low sheet resistivity, even for fast diffusing dopants such as phosphorous.

The monolayer doping (MLD) process is based on the formation of self-assembled dopant-containing monolayer on the crystalline silicon surfaces, followed by the subsequent diffusion of dopants from the surface into the lattice by a thermal annealing step (Fig. 1). In detail, for the phosphorous-MLD (P-MLD) process, 4" p-type Si wafers were first treated with dilute hydrofluoric acid (~1%) to remove the native $SiO_2$. The Si surface was then reacted with diethyl 1-propylphosphonate (DPP, Alfa Aesar) and mesitylene as a solvent (25:1, *v/v*) for 2.5 h at 120°C to assemble a P-containing monolayer. The details of this reaction and the monolayer formation kinetics have been reported elsewhere.[9,10] Then, a layer of ~50nm thick $SiO_2$ is electron-beam evaporated as a cap, and the substrate is spike annealed between 900-1050°C in Ar ambient to drive in the P atoms and achieve *n+/p* USJs. The spike annealing is performed in a rapid thermal processing tool (AG Associate, model 610) with a fast ramping rate of 100°C/s to the target temperature. During the annealing, the 4" Si wafer is placed on top of a 6" pocket wafer and temperature is monitored by the pyrometer controlling system. A similar approach was applied to 4" n-type Si wafers for boron-MLD (B-MLD) for which hydrogen terminated Si wafers were reacted with allylboronic acid pinacol ester (ABAPE, Aldrich) and mesitylene as a solvent (25:1, *v/v*) at 120ºC for 2.5 hr to enable a B-containing monolayer which is then capped with $SiO_2$ followed by spike anneal to enable the formation of *p+/n* USJs.[9] Finally, the oxide cap is removed and the enabled junctions are characterized.

Secondary ion mass spectrometry (SIMS) measurements were performed to characterize the dopant profiles. Figure 2A illustrates the phosphorous SIMS profiling for P-MLD with spike anneal temperatures of 950-1050ºC. Notably, for all samples, there is a dramatic change in the P profile at the concentration of $1\sim5\times10^{19}$ atoms/cm$^3$, which is known as the "kink-and-tail" characteristic. This behavior has been commonly observed for the conventional phosphorus doping schemes[11] and is attributed to the changeover from the vacancy assisted diffusion mechanism (at high P concentration region) to the kick-out diffusion mechanism (at low P concentration region).[12] From the temperature dependency, two trends are clearly evident. First, the surface concentration of incorporated P monotonically increases with the annealing temperature (Fig. 2B). Specifically, surface doping concentrations of $N_o \sim 2.5\times10^{20}$, $3.5\times10^{20}$, $4\times10^{20}$, $5.5\times10^{20}$ atoms/cm$^3$ are observed for 900, 950, 1000, 1050 ºC, respectively. This trend is consistent with the constant source, surface diffusion model, in which $N_o$ is governed by the dopant solubility limit at the diffusion temperature. In fact, the observed temperature dependency of $N_o$ is in close agreement with the previously reported solid solubility limits (Fig. 2B).[13,14] In the MLD process, the maximum areal dose, $Q$, corresponds to the monolayer packing density (~$8\times10^{14}$ molecules/cm$^2$ assuming a molecular footprint of ~0.12 nm$^2$); hence, at a first glance, it may seem that the limited source diffusion model is more applicable. However, in this work, since spike annealing is applied, the monolayer packing density is higher than the areal dose of the incorporated/diffused dopants. Therefore, within the context of this work, constant source model may be applied as a rough guideline in predicting the doping profile behaviors.

The second clear trend observed from the SIMS profiling is a monotonic increase in the junction depth and areal dopant dose with the diffusion temperature. The junction depth, $x_j$ is defined as the depth at which the incorporated P concentration equates the background B concentration of the substrate (~$5x10^{18}$ B atoms/$cm^3$), while $Q$ is extracted by integrating the total area of the dopant profiles. Notably, the substrate concentration of $5x10^{18}$ atoms/$cm^3$ used here is the same as the channel doping density for the state-of-the art Si MOSFETs. We extract $x_j$ ~ 2, 5, 7, 25 nm and $Q$ ~ $5.5x10^{12}$, $1.0x10^{13}$, $1.7x10^{13}$, $7.5x10^{13}$ P atoms/$cm^2$ for 900, 950, 1000, 1050ºC spike anneals, respectively (Fig. 2C). This trend is expected and arises from the enhanced diffusivity and solubility of P in Si at higher diffusion temperatures. It should be noted that given the finite temperature ramp time (~100ºC/s) of our rapid thermal annealing tool, the diffusion time is also effectively increased for the samples treated at higher annealing temperatures, which may also attribute to the observed temperature dependency of the dopant dose.

The sub-5 nm USJs with high $Q$ enabled by MLD are highly attractive, and clearly demonstrates the potency and viability of this technology for future nanoscale CMOS fabrication processing. This unique feature of MLD arises from the lack of TED during the dopant incorporation which is in distinct contrast to the ion implantation process. Additionally, in the MLD process, the incorporated dopant atoms near the surface are not lost during the $SiO_2$ cap (i.e. mask) removal step, owing to the high etch selectivity of the oxide over crystalline Si. This is in distinct contrast to the ion implantation process in which the post-implantation mask removal and surface cleaning steps lead to some dopant (and Si) loss near the surface due to the enhanced etch rate and reduced etch selectivity of the damaged (nearly amorphized by the implanted ions) top Si layer. Uniquely, this work shows that conventional annealing methods can indeed enable sub-5 nm USJs when dopants are introduced from the surface which is yet another beneficial aspect of MLD since the uniformity and reliability of sub-milli-second, non-equilibrium annealing methods[7,8] (i.e., flash and laser) are still unknown and under active investigation.

While SIMS is highly valuable for obtaining the overall dopant profiles, some uncertainty and error may be expected in the measured profiles, especially for the first 1-2 nm depths from the surface and even when the measurement tool is cautiously operated near the depth resolution limit (see Supporting Information). Additionally, SIMS does not provide information on the electrically active content of the incorporated dopants, which is critical for the device applications. Therefore, to further characterize the *n*+/*p* USJs and examine the electrically active concentration of the incorporated dopants, sheet resistance ($R_s$) measurements are imperative. Accurate $R_s$ measurements, however, are quite challenging for USJs. Specifically, conventional contact-mode, four-point probe measurements cannot be utilized for sub-10nm USJs because of the probe penetration into the surface of the substrate.[15] This probe penetration causes significant junction damage and leakage which underestimates the true $R_s$ with a corresponding error as high as 100% or more.[15] In this aspect, we utilized the non-contact $R_s$ technique to electrically characterize our USJs. Briefly, this method is relied on the principle of measuring the difference in the surface photo-voltage between two non-contact, voltage probes as induced by an external light source (Fig. 3). By varying the light modulation frequency, the spatially resolved surface voltage can yield an accurate estimation of $R_s$ and the junction leakage current density with the details reported elsewhere.[16] Figure 4 shows the non-contact $R_s$ wafer maps for the *n+/p* junctions enabled by MLD. Average $R_s$ ~ 12000, 3670, 3160, 825 Ω/□ are observed for the 900, 950, 1000 and 1050°C spike anneals, respectively. This temperature dependency is expected due to the higher diffusivity and solubility of dopants (i.e., higher $Q$) at higher temperatures, which effectively results in higher free carrier concentrations and lower $R_s$.

From the wafer-scale $R_s$ maps, a modest standard deviation of $\sigma \leq 30\%$ is obtained for all samples, except for the 900°C spike annealed wafer (Fig. 4). The large variation (σ~100%) in the $R_s$ for the 900°C spike wafer may be due to the average measured $R_s$ ~12,000 Ω/□ being close to the resolution limit of the non-contact $R_s$ measurement set up. Notably, clear rotational symmetry is observed in the $R_s$ maps for all samples with $R_s$ being highly uniform (σ<5%) within each circular "ring" boundaries. This variation pattern is a signature of the spatial imbalance in the power density of the heater lamp[17] which may be expected since the annealing tool used in this study was not designed for spike anneal applications[18]. Notably, the "ring" pattern is off-centered for the 1050°C wafer (Fig. 4D) because the wafer was misaligned from the center of the support substrate during the annealing step, again indicative of the role of the non-uniform heating in the observed $R_s$ variation. Given that $R_s$ is highly uniform within each "ring", we speculate that in the future, a higher uniformity across the wafer may be attained by MLD if a more sophisticated spike annealing tool is used. This is an expected feature of MLD since the dopants are deterministically positioned on the surface of the wafer through a self-limiting monolayer formation reaction, therefore, providing a high degree of control in the uniformity of surface dopant coverage prior to the diffusion step.

In order to compare our P-MLD processed $n+/p$ USJs with those achieved by other doping technologies, we complied the literature reported $R_s$ and $x_j$ values for phosphorus doped junctions as depicted in Fig. 5A.[19,20,21,22,23] From the literature, the smallest $x_j$~13nm at the background concentration of $5x10^{18}$ atoms/cm$^3$ was reported with $R_s$ ~650 Ω/□. To the best of our knowledge, there is no previous report of sub-10nm $n+/p$ USJs based on the phosphorous diffusion in part because of the high diffusivity of P, highlighting the elegance of MLD in achieving nm-scale junctions, even for fast diffusing impurities. Notably, for $x_j$~25nm, the $R_s$ values obtained from MLD are comparable (within a factor of ~2) with those obtained by other conventional doping methods (Fig. 5A). Moreover, a simple, analytical constant-source diffusion modeling (P in Si) was carried out (Fig. 5A) to further shed light on the MLD experimental data (see Supp. Info. for details). The experimental values qualitatively fit the modeling trend, again demonstrating the near ideal behavior of the MLD process.

From the SIMS and $R_s$ measurements, the phosphorous activation efficiency, η, was directly obtained for each diffusion temperature (Fig. 5B) in order to shed light on the percentage of the incorporated dopants that are electrically active. Specifically, $R_s$ was estimated from the SIMS profiles by equation 1 and then compared to the non-contact $R_s$ measurement values,

$$R_{s,SIMS}{}^{-1} = \int_0^{x_j} q\mu(N)N(x)dx \qquad \text{Eq. 1}$$

where $q$, $\mu$, $N$ and $x$ are the elemental charge, electron mobility, dopant concentration and depth, respectively. The efficiency is then defined as $\eta = \frac{R_{s,SIMS}}{R_{s,measurement}}$. Figure 5B shows the extracted efficiencies of η~70% for spike annealing temperatures ≥ 950°C with η~30% for the 900 °C spike anneal. The discrepancy between the measured and calculated Rs values may be attributed to some percentage of the diffused dopants that are incorporated in sites other than substitutional lattice sites.[12] However, since in the MLD, dopants are introduced from the surface by an equilibrium process, we expect the diffusion of dopants to be dominated by the substitutional mechanism, resulting in nearly all the incorporated dopants being electrically active. Therefore, we speculate that the small discrepancy between the measured and calculated $R_s$ values arise from the uncertainty in the SIMS profiles. This error is particularly magnified for the 900$^o$C wafer since the observed junction depth is only ~2 nm for this sample, right at the resolution limit of SIMS. The high η estimated from the data analysis is yet another highly attractive feature of MLD since for device applications, only the electrically active contents are desirable with other dopants inducing defects and/or junction leakage currents.

Since organic molecular precursors are utilized as the dopant source, carbon incorporation during the dopant diffusion may be expected, which may enhance the highly undesired junction leakage currents. Therefore, investigations of C incorporation and the arising junction properties are needed. However, SIMS measurements do not provide accurate C depth profiling near the surface region due to the atmospheric organic surface which is unavoidable even if a pre-measurement cleaning step is performed on the samples.[24] Instead, in order to directly characterize the effect of potential carbon incorporation, the junction leakage current density was measured by the non-contact, photo-voltage measurement. Figure S3 illustrates the leakage wafer map for the *n+/p* junctions for various spike temperatures. The average leakage currents of ~0.13, 0.55, 0.11, 0.31 μA/cm$^2$ were measured for the 900, 950, 1000, 1050 $^o$C spike anneals, respectively. Notably, the small leakage currents (<1 μA/cm$^2$) are close to the resolution limit of the instrument, and attests to the high quality junctions that are enabled by MLD. The results suggest that carbon content incorporation may not be of a concern for MLD, at least when considering the junction leakage currents, which are smaller than the start-of-the-art USJs' leakage currents.[25]

In addition to the n+/p junctions, we also investigated the formation of *p+/n* USJs by boron-MLD with spike annealing. From SIMS measurements, sub-2 nm junction depths ($x_j$ ~ 1 and 2 nm) are obtained for 950 and 1050$^{o}$C spike, respectively (Fig. S1), which is close to the resolution limit of SIMS. The shallower junctions enabled at the same diffusion temperature for B-MLD as compared to P-MLD is expected due to the lower diffusivity of B. Since the B-MLD p+/n junctions are at the molecular-scale, the $R_s$ values were higher than ~10,000 Ω/□ for all spike conditions, and therefore, out of the measurement range of the non-contact photo-voltage characterization technique. In the future, more advanced electrical and materials characterization methods, such as local electrode atom probe microscopy need to be utilized to further study the amazingly shallow *p+/n* USJs enabled by MLD.

In summary, we have demonstrated the wafer-scale formation of *n+/p* and *p+/n* USJs by the combination of self-limiting monolayer doping and the conventional spike annealing. This approach is demonstrated on 4" Si wafers with the junction uniformity limited by the temperature homogeneity of the spike anneal tool. For phosphorous doping, we report for the first time, sub-10 nm junction formation (down to 2nm – the SIMS resolution limit) with the non-contact $R_s$ measurements being consistent with the predicted values from the dopant profiles. Additionally, we find minimal junction leakage currents which are indicative of high quality, defect-free USJs enabled by MLD. Notably, besides nano-scale controlled doping for the contact extension of future MOSFETs, this surface doping technology also may be highly applicable for the conformal and deterministic doping of non-planar nanoscale device structures, such as Fin-FETs or nanowire-FETs.

**Acknowledgements**

This work was financially supported by SEMATECH, NSF, Intel Corporation, and MARCO MSD Focus Center Research Program. J.C.H acknowledges a graduate student fellowship from Intel Foundation.

**Supporting Information Available:**

Details of SIMS measurements; activation efficiency (η) calculations; constant source diffusion modeling; non-contact junction leakage measurements. These materials are available free of charge via the Internet at http://pubs.acs.org.

**Figure Captions:**

**Figure 1.** Process schematic for the wafer-scale monolayer doping approach.

**Figure 2.** Phosphorus monolayer doping characterization. (A) Secondary Ion Mass Spectrometry (SIMS) profile of phosphorus atoms for different spike anneal temperatures. (B) Phosphorus surface concentration, $N_o$ obtained from SIMS analysis for MLD processed samples as a function of spike annealing temperature. For comparison the previously reported solid-solubility limits for different temperatures are also shown. (C) Phosphorus areal dose versus junction depth (at a background of $5x10^{18}$ atoms/cm$^3$) for different spike anneal temperatures.

**Figure 3.** Non-contact Rs and junction leakage currents measurement schematic.

**Figure 4.** Non-contact sheet resistance (in Ω/□) wafer map for P-MLD samples with spike anneal temperatures of (A) $900^0$C, (B) $950^0$C, (C) $1000^0$C and (D) $1050^0$C.

**Figure 5.** (A) Sheet resistance (in Ω/□) versus $x_j$ for phosphorus doped Si samples reported in this work (by P-MLD) and the literature (by conventional doping methods). The dot line shows a simple constant source diffusion model (see Supp. Info.) for comparison purposes. (B) Dopant activation efficiency for P-MLD samples as a function of spike annealing temperature.

**Reference**

[1] Peercy, P. S. The drive to miniaturization. *Nature*, **2000**, 406, 1023-1026.

[2] Claeys, C. Technological challenges of advanced CMOS processing and their impact on design aspects. *VLSI Design*, **2004**, 2004, 275-282.

[3] Stolk, P.A.; Gossmann, H.J.; Eaglesham, D.J.; Poate, J.M. Implantation and transient boron diffusion: the role of Si self interstitials. *Nucl. Instrum. Methods Phys. Res. B,* **1995**, 96, 187.

[4] Kawasaki,Y.; Kuroi T.; Yamashita T.; Horita K.; Hayashi T.; Ishibashi M.; Togawa M.; Ohno Y.; Yoneda M.; Horshy T.; Jacobson D.; Krull W. Ultra-shallow junction formation by B18H22 ion implantation. *Nuclear Instruments and Methods in Physics Research B,* ***2005,*** 237, 25-29.

[5] MacCrimmon, R.; Hautala, J.; Gwinn, M.; Sherman, S. Gas cluster ion beam infusion processing of semiconductors. *Nuclear Instruments and Methods in Physics Research B*, **2006**, 242, 427-30.

[6] Yon, G.H.; Buh, G.H.; Park, T.; Hong, S.J.; Shin, Y.G.; Chung, U.; Moon, J.T.; Ultra Shallow Junction Formation Using Plasma Doping and Laser Annealing for Sub-65 nm Technology Nodes. Jpn. J. Appl. Phys. **2006**, 45, 2961-2964.

[7] Ito, T.; Iinuma, T.; Murakoshi, A.; Akutsu, H.; Suguro, K.; Arikado, T.; Okumura, K.; Yoshioka, M.; Owada, T.; Imaoka, Y.; Murayama, H.; Kusuda, T. 10–15 nm Ultrashallow Junction Formation by Flash-Lamp Annealing. Jpn. J. Appl. Phys. **2002**, 41, 2394-98.

[8] Poon, C.H.; Cho, B.J.; Lu. Y.F.; Bhat, M.; See, A. Multiple-pulse laser annealing of preamorphized silicon for ultrashallow boron junction formation. *J. Vac. Sci. Technol. B.* **2003**, 21 (2), 706-709.

[9] Ho, J.C.; Yerushalmi, R.; Jacobson, Z.A.; Fan, Z.; Alley, R.L.; Javey, A. Controlled nanoscale doping of semiconductors via molecular monolayers. *Nature Materials*, **2008**, 7 (1), 62-67.

[10] Yerushalmi, R.; Ho, J.C.; Fan, Z.; Javey, A. Phosphine Oxide Monolayers on $SiO_2$ Surfaces. *Angew. Chem. Int. Ed.* **2008**, 47, 4440-4442.

[11] Bentzen, A.; Schubert, G.; Christensen, J.S.; Svensson, B.G.; Holt, A. Influence of temperature during phosphorus emitter diffusion from a spray-on source in multicrystalline silicon solar cell processing. *Prog. Photovolt: Res. Appl.* **2007**, 15, 281–289.

[12] Uematsu, M. Simulation of boron, phosphorus, and arsenic diffusion in silicon based on an integrated diffusion model, and the anomalous phosphorus diffusion mechanism. *Journal of Applied Physics,* **1997***,* 82-5, 2229.

[13] Trumbore, F.A.; Solid Solubilities of Impurity elements in Ge and Si. *Bell Syst. Tech. J.*, **1960**, 35, 205.

[14] Borisenko, V.E.; Yudin, S.G. Steady-State Solubilities of Subsitutional Impurities in Si. *Phys. Stat. Solidi*, **1987**, A (101), 1, 123.

[15] Clarysse, T.; Vanhaeren, D.; Vandervorst, W. Impact of probe penetration on the electrical characterization of sub-50 nm profiles. *J. Vac. Sci. Technol. B*, **2002**, 20 (1), 459-66, 2002.

[16] Faifer, V.N.; Current, M.I.; Wong, T.M.H.; Souchkov, V.V. Noncontact sheet resistance and leakage current mapping for ultra-shallow junctions. *J. Vac. Sci. Technol. B.,* **2008**, 26 (1), 420-24.

[17] Faifer, V.N.; Current, M.L.; Nguyen, T.; Wong, T.M.H.; Souchkov, V.V. Non-contact measurement of sheet resistance and leakage current: applications for USJ-SDE/halo junctions. *Ext. Abs. of the* $5^{th}$ *IWJT*, IEEE, **2005**, 45-8.

[18] Since the 4" wafer is placed on a support substrate during the spike annealing (the tool was designed to handle 6" wafers), this may also contribute to the variation pattern.

[19] Collarta, E. J.H.; Felch, S.B.; Pawlak, B.J.; Absil, P.P.; Severi, S.; Janssens, T.; Vandervorst, W. Co-implantation with conventional spike anneal solutions for 45 nm *n*-type metal-oxide-semiconductor ultra-shallow junction formation. *J. Vac. Sci. Technol. B.* **2006,** 24(1)**,** 507-9.

[20] Augendre E, Pawlak B.J., Kubicek S., Hoffmann T., Chiarella T., Kerner C., Severi S., Falepin A., Ramos, J.; De Keersgieter, A.; Eyben, P.; Vanhaeren, D.; Vandervorst ,W.; Jurczak, M.; Absil, P.; Biesemans, S. Superior N- and P-MOSFET scalability using carbon co-implantation and spike annealing. *Solid State Electronics*, **2007**, 51 (11-12), 1432-6.

[21] Lee, S.W.; Jeong, J.Y.; Park, C.S.; Kim, J.H.; Ji, J.Y.; Choi, J.Y.; Lee, Y.J.; Han, S.H.; Kim, K.M.; Lee, W.J.; Rha, S.K.; Oh, J.K. The study of plasma doping process for ultra shallow junctions. *Ext. Abs. of the 7th IWJT*, IEEE, **2007**, 67-8.

[22] Cagnat, N.; Laviron, C.; Mathiot, D.; Rando, C.; Juhel, M. Shallow Junction Engineering by Phosphorus and Carbon Co-implantation: Optimization of Carbon Dose and Energy. *Mater. Res. Soc. Symp. Proc.* **2007,** 994, 0994-F08-04.

[23] Pawlak, B.J.; Duffy, R.; Augendre, E.; Severi, S.; Janssens, T.; Absil, P.; Vandervorst, W.; Collart, E.; Felch, S.; Schreutelkamp, R.; Cowern, N. The Carbon Co-Implant with Spike RTA Solution for Phosphorus Extension. *Mater. Res. Soc. Symp. Proc.* **2006, 912,** 0912-C01-06.

[24] Beebe, M.; Bennett, J.; Barnett, J.; Berlin, A.; Yoshinaka, T. Quantifying residual and surface carbon using polyencapsulation SIMS. *Applied Surface Science*. **2004**, 231-32, 716-19.

[25] Chen, J.T.C.; Dimitrova, T.; Dimitrov, D. A New Method for Mapping Ultra-Shallow Junction Leakage Currents. *International Workshop on IWJT*, IEEE, **2006.** 100.

# Figure 1

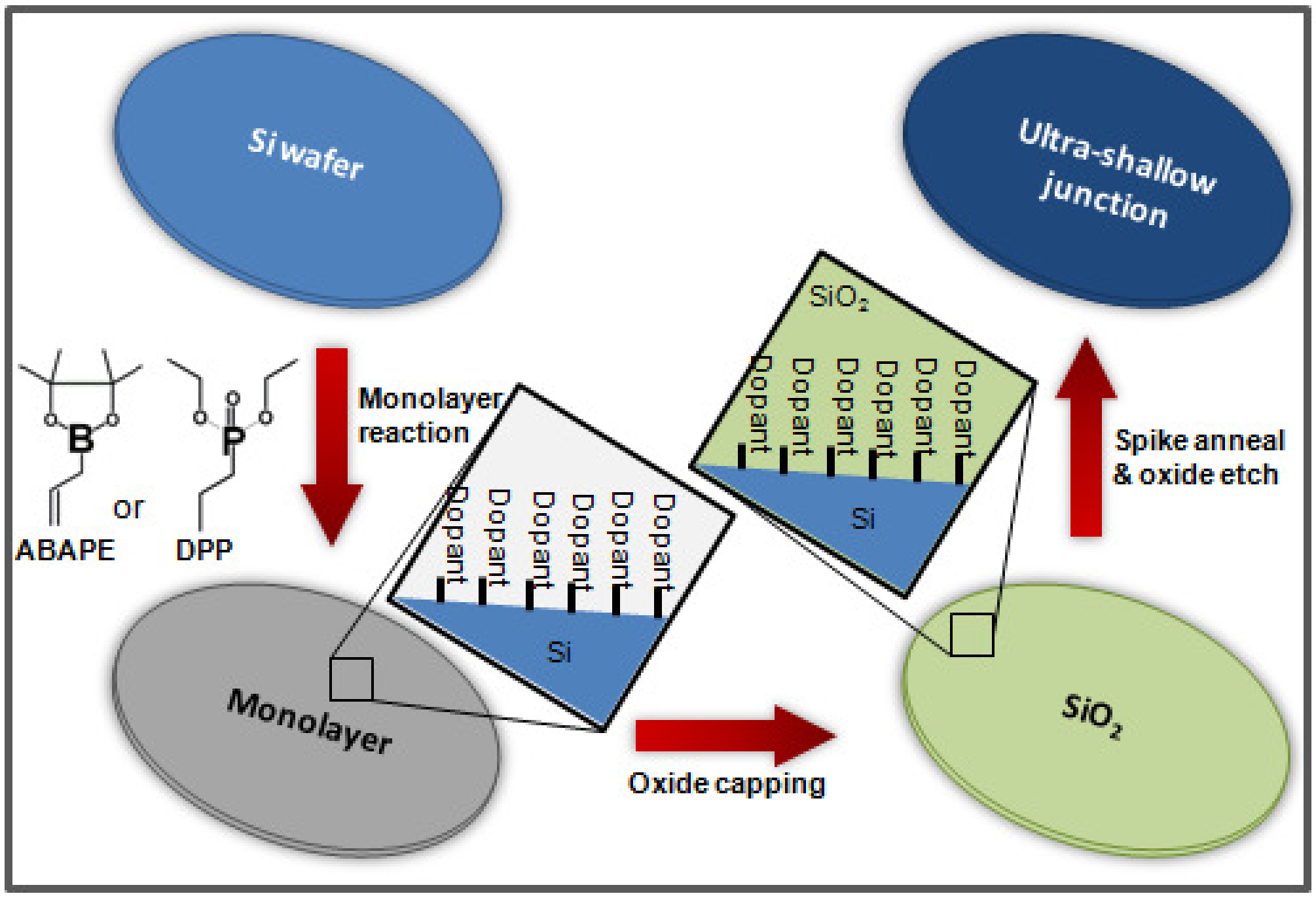

# Figure 2

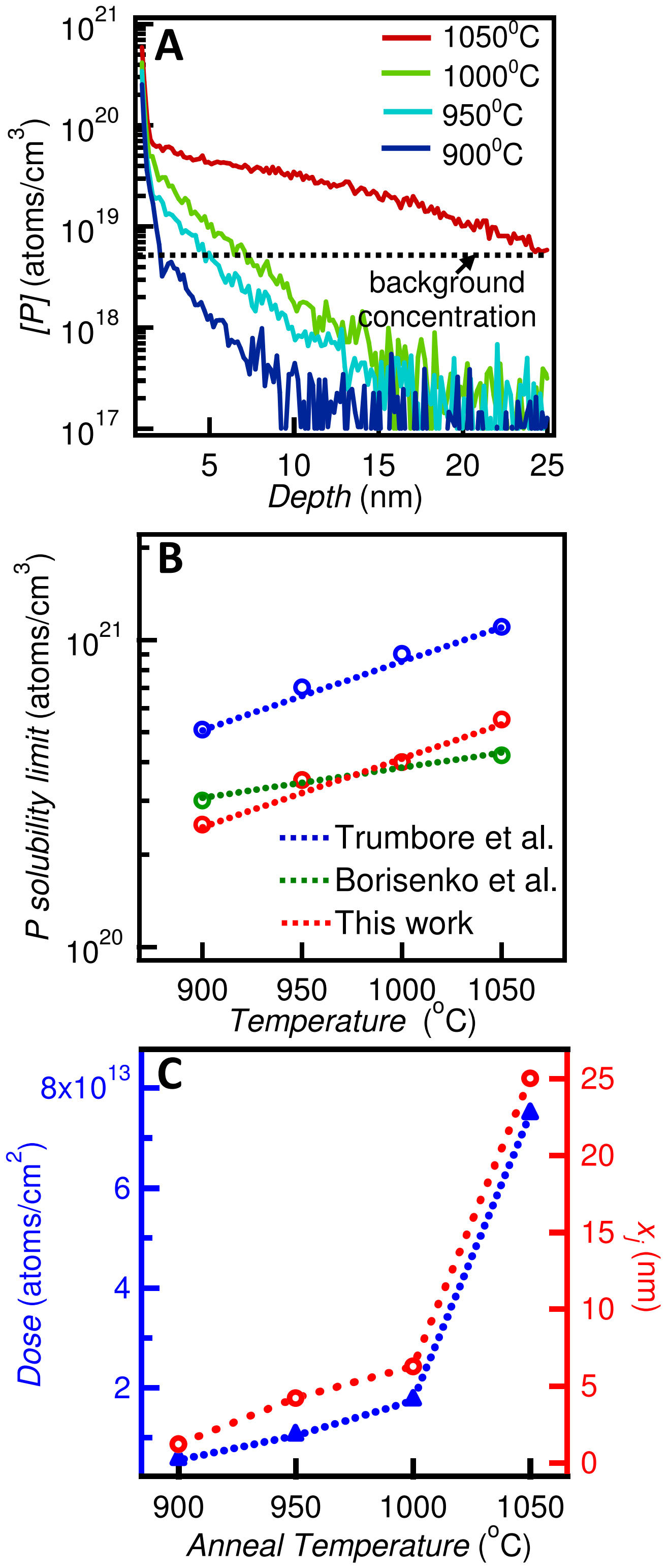

# Figure 3

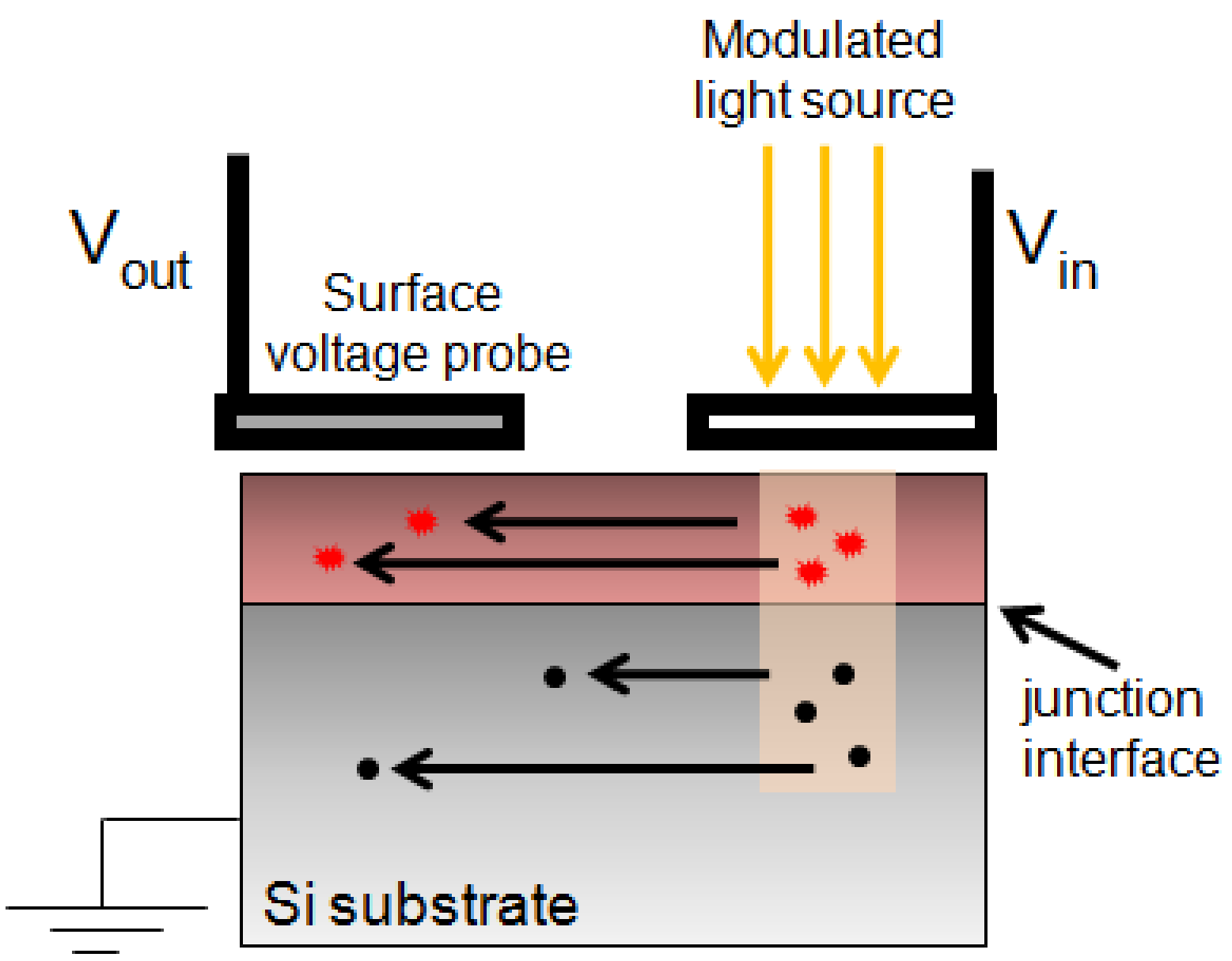

# Figure 4

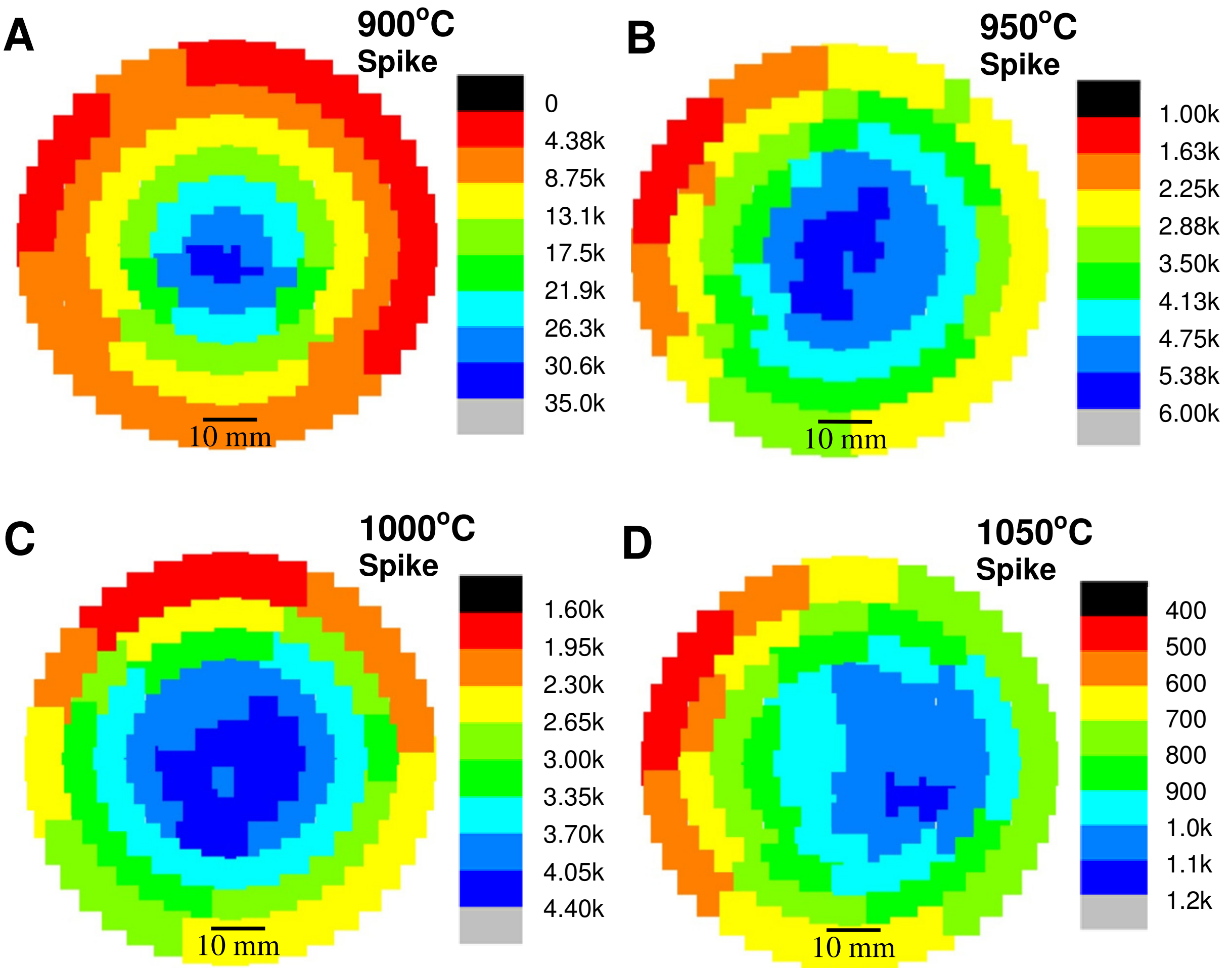

## Figure 5

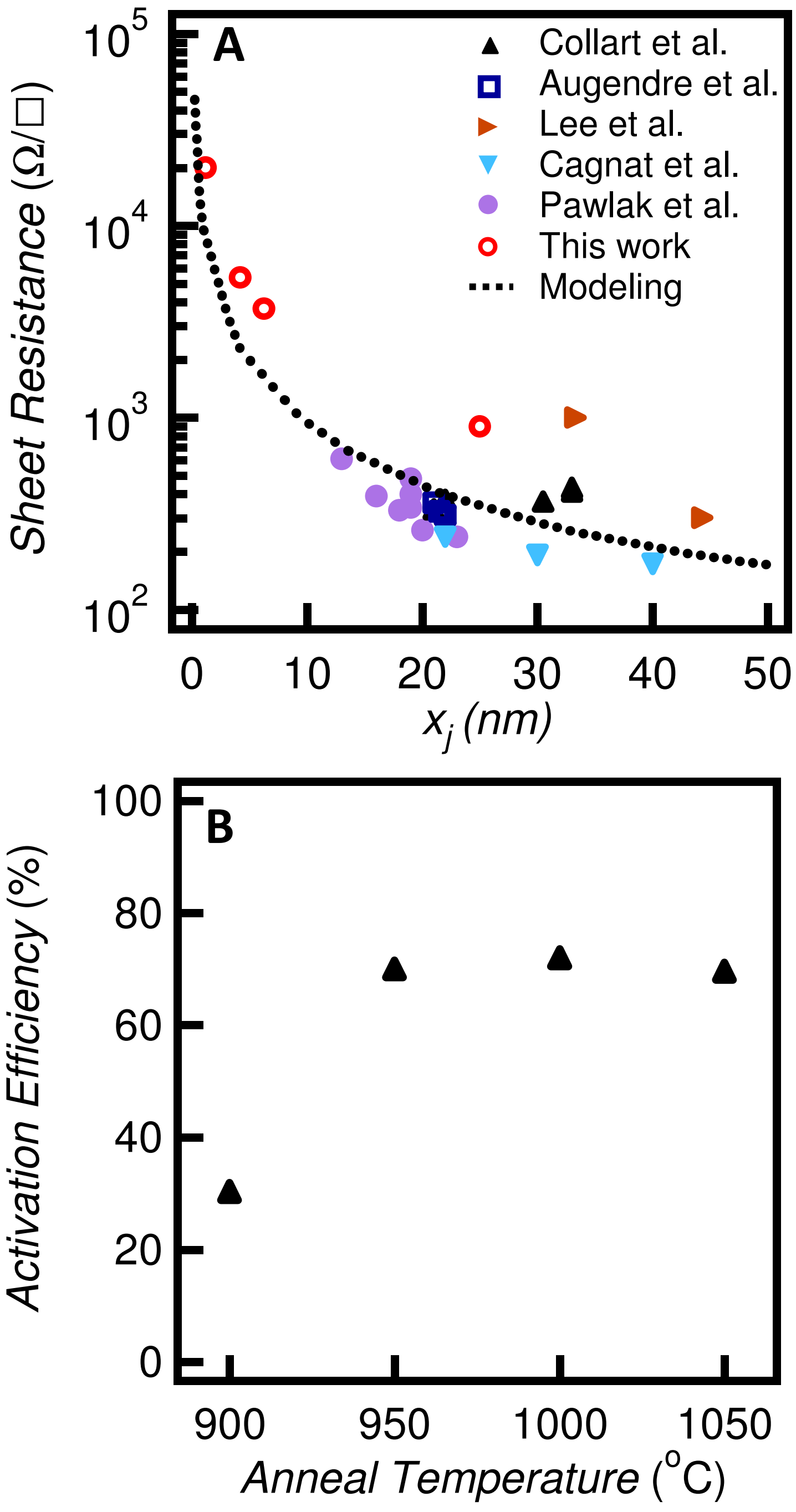

# TOC Figure

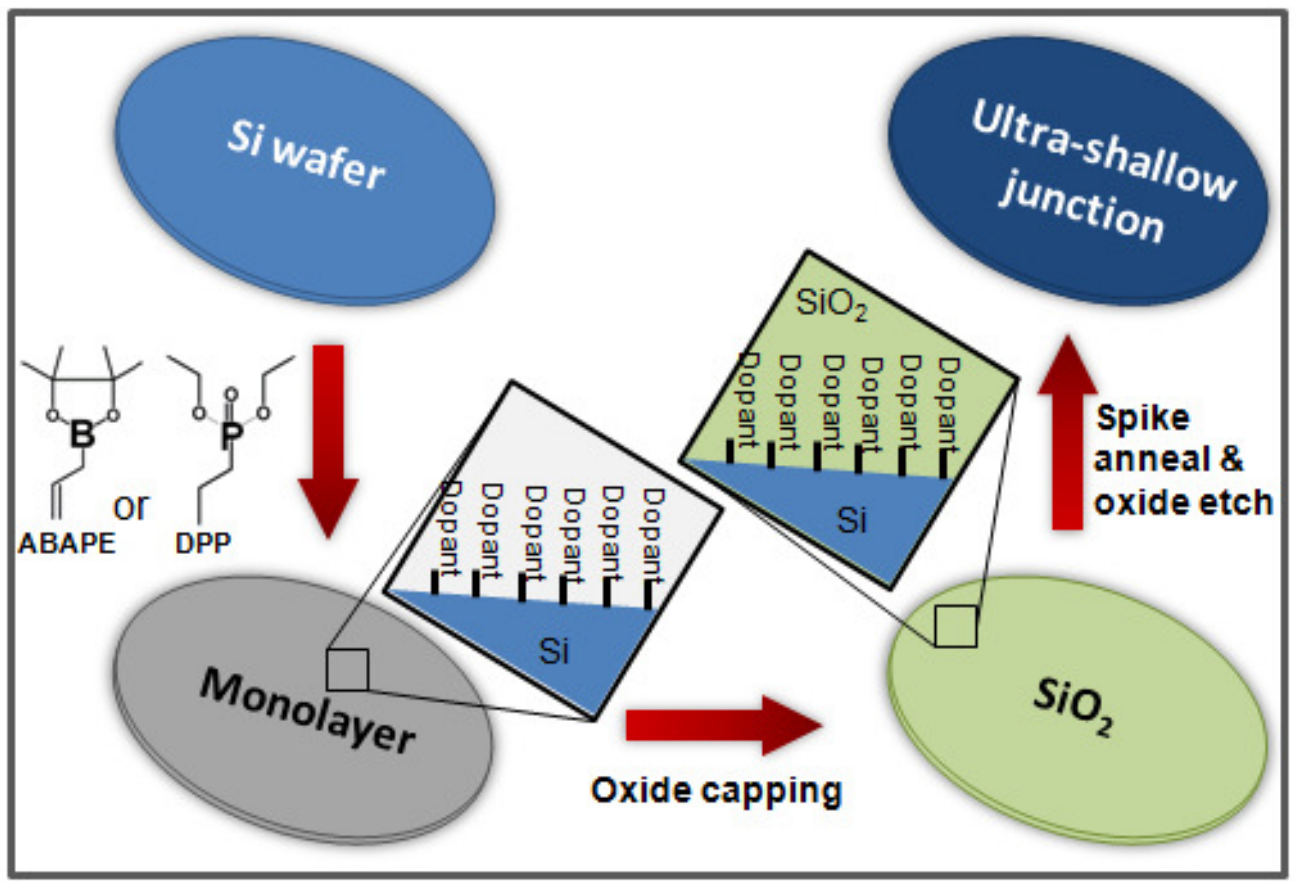

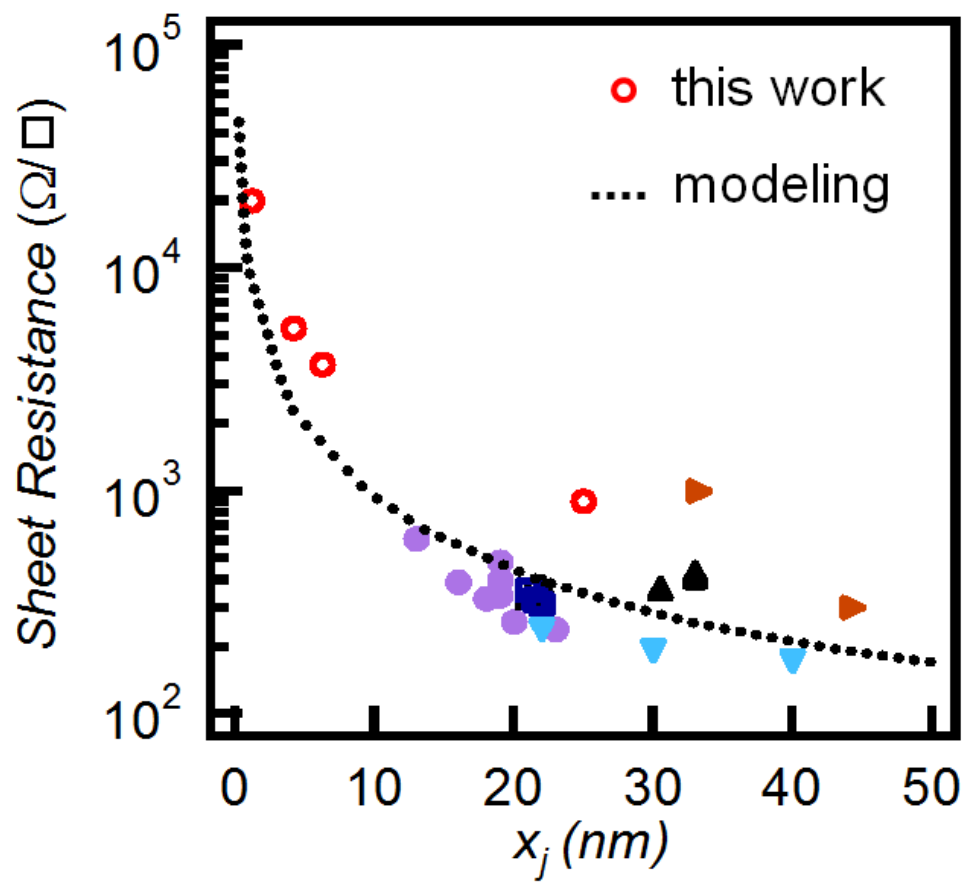